# Indexing by Latent Dirichlet Allocation and Ensemble Model[†]

Yanshan Wang, In-Chan Choi*
School of Industrial Management Engineering, Korea University
Seongbuk-gu, Seoul, Republic of Korea

Jae-Sung Lee
Diquest, #501, Kolon Villant 2
Guro-gu, Seoul, Republic of Korea

## Abstract

The contribution of this paper is two-fold. First, we present Indexing by Latent Dirichlet Allocation (LDI), an automatic document indexing method. The probability distributions in LDI utilize those in Latent Dirichlet Allocation (LDA), a generative topic model that has been previously used in applications for document retrieval tasks. However, the ad hoc applications, or their variants with smoothing techniques as prompted by previous studies in LDA-based language modeling, result in unsatisfactory performance as the document representations do not accurately reflect concept space. To improve performance, we introduce a new definition of document probability vectors in the context of LDA and present a novel scheme for automatic document indexing based on LDA. Second, we propose an Ensemble Model (EnM) for document retrieval. The EnM combines basis indexing models by assigning different weights and attempts to uncover the optimal weights to maximize the Mean Average Precision (MAP). To solve the optimization problem, we propose an algorithm, EnM.B, which is derived based on the boosting method. The results of our computational experiments on benchmark data sets indicate that both the proposed approaches are viable options for document retrieval.

Introduction

With the continuous growth in the size of information sources available on the web, Document Indexing (DI) is gaining attention as a crucial technique to retrieve significant information for users (Choi & Lee, 2010). Using DI, text documents are converted into *index terms* or *document features* that can be trivially analyzed by computers. With an appropriate *ranking function* or *retrieval model*, DI has been shown to be effective for document retrieval (Croft, Metzler, & Strohman, 2010).

Exact keyword matching is one of the earliest and simplest methods applied to DI. In retrieval models (such as the Boolean model) that use exact keyword matching, documents that contain each index term in the query are retrieved from the document set. Users are provided with these documents for seeking the desired information. The drawback of exact keyword matching is that a partial match or ranking of the retrieved documents is not considered, which often leads to poor retrieval results.

In order to overcome the aforementioned drawback, Vector Space Model (VSM) was proposed to represent a document as a vector of index term weights so that a continuous degree of relevant documents can be retrieved (Salton, Wong, & Yang, 1975; Caid, Dumais, & Gallant, 1995; Salton & McGill, 1986). In general, VSM-based retrieval models comprise three stages: (a) removing nonsignificant words from the documents using a stop list; (b) weighting the indexed terms into a vector space; and (c) ranking the documents with respect to the input query according to different similarity measures. Many variants of the VSM have been developed by modifying the weighting scheme in the second stage. One of the most well-known weighting schemes is the *tf-idf* scheme (Salton & McGill, 1986). In this scheme, each document (as well as each query) is represented by a fixed vector where each component is the *tf-idf* term weight. One of the drawbacks in the *tf-idf* scheme is that it ignores the conceptual meaning of words, and therefore, it suffers from difficulties associated with synonymy and polysemy (Salton & McGill, 1986). In addition to the conventional *tf-idf* scheme, Okapi BM25 is used to estimate term importance (Robertson et al., 1995). However, Okapi BM25 also ignores the interrelationship between terms within a document (Büttcher, Clarke, & Lushman, 2006).

Document Modeling (DM) is the task of finding inherent structures and features in documents. It provides a new means for DI because document features can be regarded as index terms. Among the DM methods, a concept model analyzes documents in a concept space by considering a hidden layer of the interweaving relationship between terms, which allows it to overcome the difficulties of the VSM using concept representation for documents. The concept model assumes that terms appearing frequently in documents are likely to be related to each other through unidentified concepts (Guha, McCool, & Miller, 2003; Nikravesh, 2007; Gunter, 2009; Trillo, Po, Ilarri, Bergamaschi, & Mena, 2011). Latent Semantic Analysis (LSA) is a successful deterministic approach in concept modeling (Deerwester, Dumais, Landauer, Furnas, & Harshman, 1990). LSA utilizes Singular Value Decomposition (SVD) and projects high-dimensional data into a lower dimensional space to overcome the overfitting problem. In addition, LSA captures some aspects of synonymy and polysemy by deriving conceptual features of original *tf-idf* weights. These conceptual features can be utilized for automatic indexing and retrieval. Despite the merits of the LSA, however, it is difficult to interpret the outputs obtained from the analysis, especially when the output values are negative.

Probabilistic concept model is a probabilistic approach to model documents in terms of latent concepts. Indexing and retrieval using probabilistic concept models are based on the assumption that the concepts are distributed differently in relevant and non-relevant documents. Probabilistic Latent Semantic Analysis (pLSA) (Hofmann, 2001), a generative probabilistic concept model based on LSA, provides clarity in the interpretation of output values since they have meanings in terms of probability. Probabilistic Latent Semantic Indexing (pLSI) is an application of the pLSA in automatic indexing. pLSI has a limitation in that it does not explicitly show how to assign probabilities to documents that are not in the training set, and that it has to deal with the difficulties of parameter interpretation. Further, pLSI always overfits training data sets when the number of parameters increase with the increasing number of documents in a corpus.

Latent Dirichlet Allocation (LDA) is an alternative generative probabilistic concept model (Blei, Ng, & Jordan, 2003). By importing a symmetric Dirichlet prior, the LDA resolves the limitations associated with pLSI (Wallach, Mimno, & McCallum, 2009). LDA addresses the topic-based structural analysis of corpora, and thus it can be regarded as a model for topic search. LDA is mainly used in DM and classification and only a few research studies, such as (X. Wei & Croft, 2006), apply LDA in the context of query searches. These studies mainly focus on the methods of avoiding the assignment of a zero value to the conditional probability of a query given a document. For example, the use of gamma values in the LDA, or some forms of smoothing models have been proposed (Azzopardi, Girolami, & Van Rijsbergen, 2004). However, it is still unclear how to apply LDA to automatic indexing for the query search.

A ranking function is used in conjunction with DI to calculate the relevance between a document and an input query. In general, the degree of relevance between the document and the query are determined using similarity measures. There are many approaches to measure similarity, such as relevance scores (Singhal, 2001), however, most of approaches are empirical functions based on experimental evaluations. A widely used similarity measure is cosine similarity, which calculates the cosine value of the angle between the query vector and the document vector (Salton & McGill, 1986). The advantage of the cosine similarity measure is that it considers two documents as identical if they have the same proportion in term frequencies regardless of the number of terms. The cosine similarity measure is generally utilized with indexing methods for examining retrieval performance (Deerwester et al., 1990; Hofmann, 2001). Since we intend to compare the retrieval performances of different indexing methods, we employ the cosine similarity measure in our approach as well.

Unlike the retrieval model that integrates an indexing method and a ranking function, an ensemble model (EnM) is a discriminative retrieval model based on a linear combination of a number of basis models. The EnM ranks documents according to the summation of weighted similarity values that are computed by constituent indexing models. However, different weights assigned to the constituent models result in distinct performances for document retrieval. The optimization problem of calculating weights of an EnM has been well studied in the classification domain whereas it is rarely tackled in the information retrieval domain. AdaRank (Xu & Li, 2007) is by far the only well known learning algorithm that uses the AdaBoost framework (Freund & Schapire, 1995). AdaRank repeatedly constructs weak rankers and finally linearly combines them into a strong ranker. The disadvantage of AdaRank is that the number of iterations is difficult to decide. Thus, a sufficiently large human determined constant is assigned to the number of iterations. iRANK (F. Wei, Li, & Liu, 2010) is also a recently proposed ensemble model; however, it can only combine two rankers.

Our contribution is divided in two parts in this paper. In the first part, we propose an LDA-based probabilistic topic search model that identifies a set of documents that closely matches a given set of query terms in a topic space. This new method is called *Indexing by Latent Dirichlet Allocation* (LDI). In this method, we use LDA to analyze the document structure according to the hidden topics, and index documents in terms of topics by novelly defining document probability vectors in topic space. Then, we use these vectors to calculate the similarity between documents and queries. In doing so, we can identify the conceptual relevant documents given an input query, and thus, migrate the common linguistic phenomena, synonymy and polysemy.

In the second part, we propose an EnM that directly maximizes the Mean Average Precision (MAP) in order to obtain the optimal weights assigned to the constituent models. For solving the optimization problem, we propose an algorithm, namely EnM.B, based on the boosting scheme. The differences between EnM.B and AdaRank are that the number of iterations is not predetermined, and the constituent models are not randomly generated during iterations in our algorithm. Overall, the discriminative EnM performs better than the generic LDI. However, the EnM suffers from the disadvantages of the discriminative model. That is, the discriminative model needs training examples that may not exist in a data set. Therefore, we suggest that, (1) in systems that contain labeled items, such as library systems, the EnM is an optional method; and (2) in other systems that do not contain labeled items, such as search engines, LDI is a viable choice.

The remainder of this paper is organized as follows. In the next section, the background of LDA is presented. We then demonstrate the LDI and EnM models. Computational results of the proposed methods on publicly available data sets are subsequently reported. Some concluding remarks are stated in the last section.

## Latent Dirichlet Allocation

The LDA model is based on the assumptions of bag-of-words that the order of words in a document can be neglected, and all documents in a corpus share a number of latent topics. Under this assumption, a collection of documents can be represented by a collection of topics, and each document can be represented by the mixtures of these latent topics in specific proportions.

DA is a typical directed probabilistic graphical model. It has a clear three-level structure: corpus level, document level, and word level. Each level is presented by corresponding parameters and random variables. The process of generating words for each document is illustrated as follows (Blei, 2012):

1. Randomly choose a proportion over topics.
2. For each word in the document,

- Randomly choose a topic from the proportion over topics in step 1.
- Randomly choose a word from the corresponding proportion over the vocabulary.

In the following, we use notations to mathematically illustrate the LDA model. A document $\mathbf{d}$ is defined as a sequence of $N$ words, i.e., $\mathbf{d} = (w_1, w_2, ..., w_N)$, where $w_n$ denotes the $nth$ word in the sequence, and a corpus $C$ is defined as a collection of $M$ documents, $C = \{d_1, d_2, ..., d_M\}$. From the given corpus, the LDA generates hidden topics that are obtained by inferring the topic mixture $\theta = \{\theta_1, \theta_2, ..., \theta_K\}$ in the document level. Further, a set of $N$ topics in the word level is defined as $z = (z_1, z_2, ..., z_N)$. In addition, a vocabulary index set $\{1, 2, ..., V\}$ is maintained to indicate whether a particular word is used or not, i.e., $w^j = 1$ if the $jth$ word in the vocabulary list is used and $w^j = 0$, otherwise. Note that we will also use $w^j$, when used alone unaccompanied by the equality, to denote $jth$ word in the vocabulary list in other sections.

Mathematically, the process of LDA to generate a document consists of three concrete steps (Blei et al., 2003):

1. Choose the number of words $N \sim$ Poisson($\xi$).
2. Choose $\theta \sim$ Dirichlet($\alpha$).
3. For $n = 1, 2, ..., N$

- Choose a topic $z_n \sim$ Multinomial($\theta$);
- Choose a word $w_n \sim$ Multinomial($w_n|z_n, \beta$), a multinomial distribution conditioned on the topic $z_n$.

In the above steps, Dirichlet parameter $\alpha \in \mathbb{R}_+^K$, multinomial parameter $\theta \in \mathbb{R}^K$, $\theta_k \geq 0$ and $\sum_k^K \theta_k = 1$, and the corresponding dimension $K$ are all assumed to be known. The conditional probability of $jth$ word in the vocabulary list, given that the $kth$ topic is selected, is denoted by $\beta_{kj} = p(w^j = 1|z^k = 1)$. Its maximum likelihood estimator can be obtained from a posterior probability distribution. The matrix of the conditional probabilities is denoted by $\beta = [\beta_{kj}] \in \mathbb{R}^{K \times N}$. The joint prior probability distribution $p(\theta, \mathbf{z}, \mathbf{d}|\alpha, \beta)$ is expressed as

$$p(\theta, \mathbf{z}, \mathbf{d}|\alpha, \beta) = p(\theta|\alpha) \prod_{n=1}^{N} p(z_n|\theta) p(w_n|z_n, \beta), \tag{1}$$

where $p(z_n|\theta) = \theta_k$ for the unique $k$ such that $z_n^k = 1$. Here, $z_n^k$ is a binary variable indicating whether the $kth$ topic is used in selecting the $nth$ word in the document. We note that the superscript represents the order in which a word or a topic appears in the vocabulary list and the topic list, and the subscript denotes the order in which a word or a topic appears in a document.

Introducing superscript representation makes it easier to derive the marginal probability of the word appearance obtained by integrating over $\theta$ and summing over $\mathbf{z}$ on the prior, i.e.,

$$\begin{aligned} p(\mathbf{d}|\alpha, \beta) &= \int \sum_{\{z_n\}} p(\theta, \mathbf{z}, \mathbf{d}|\alpha, \beta) d\theta \\ &= \int p(\theta|\alpha) \sum_{\{z_n\}} \prod_{n=1}^{N} p(z_n|\theta) p(w_n|z_n, \beta) d\theta \\ &= \int \prod_{n=1}^{N} \sum_{z_n} p(z_n|\theta) p(w_n|z_n, \beta) d\theta. \end{aligned} \tag{2}$$

Hence, the corpus probability is obtained as

$$p(C|\alpha,\beta) = \prod_{d=1}^{M} \int p(\theta_d|\alpha) \prod_{i=1}^{N_d} \sum_{z_{dn}} p(z_{dn}|\theta_d) p(w_{dn}|z_{dn},\beta) d\theta_d, \quad (3)$$

where $\theta_d$'s are variables in the document level and $w_{dn}$ and $z_{dn}$, sampled once for each word in each document, are variables in the word level.

Using equations (1) and (2), we can specify the posterior distribution of the hidden variables $\theta$ and $\mathbf{z}$ as

$$p(\theta, \mathbf{z}|\mathbf{d},\alpha,\beta) = \frac{p(\theta,\mathbf{z},\mathbf{d}|\alpha,\beta)}{p(\mathbf{d}|\alpha,\beta)}. \quad (4)$$

Owing to the coupling relationship between variables $\theta$ and $\beta$, the maximum likelihood estimators of the posterior distribution are not tractable. To overcome this difficulty, (Blei et al., 2003) introduced free variational parameters $\gamma$ and $\varphi$ for Dirichlet and multinomial distribution, respectively, and defined variational distribution

$$q(\theta,\mathbf{z}|\gamma,\varphi) = q(\theta|\gamma) \prod_{n=1}^{N} q(z_n|\varphi_n). \quad (5)$$

The maximum likelihood estimators of $\alpha$ and $\beta$ are calculated by the EM algorithm using Jensen's inequality to estimate the lower bound on the log-likelihood of the variational distribution $q(\theta,\mathbf{z}|\gamma,\varphi)$.

## Indexing by LDA

In essence, LDA has been utilized in retrieval models such as the *query likelihood retrieval model* for ranking and retrieving documents. Azzopardi et al. (2004) first utilized one of the variational parameters $\gamma$ in LDA in Bayes smoothing and Jelinek-Mercer smoothing document retrieval models for ad-hoc retrieval. However, $\gamma$ in the LDA is auxiliary and it may not be appropriate for direct use in the query search (X. Wei & Croft, 2006). X. Wei and Croft (2006) proposed the LDA-Based Document Model (LBDM) for retrieval tasks by combining the Dirichlet smoothing method and the LDA posterior estimates of $\theta$ and $\varphi$. However, the settings for LDA estimation are based on training collection, which is coarse in the context of information retrieval. Having said that, these applications have shown the effectiveness of LDA for the retrieval task. However, LDA has been rarely applied to automatic document indexing, to the best of our knowledge. Since LDA models documents as a mixture of topics, it provides a new approach for representing documents in a topic space where the topics can be seen as index terms for indexing. In this section, we first define new term and document representations in the topic space for indexing documents and second, we demonstrate how these novel representations can be applied for the document retrieval task.

*Document Representation in Topic Space*

Since we aim to construct explicit document representations associated with topics, our method directly uses the $\beta$ matrix in the LDA model. The conditional probability $\beta_{jk}$ in LDA represents the selection probability of the word $w^j$ given a topic (concept) $z^k$. This value represents the probability of a word given a specific topic and it is used in identifying words that are associated with a topic. However, it may not be used as the probability of a topic given a word. Thus, for characterization, we define word representation in topic space, $W_j \in \mathbb{R}^K$. The *kth* component $W_j^k$ of $W_j$ represents the probability of word $w^j$ embodying the *kth* concept $z^k$. This quantity can be obtained by Bayes' rule as

$$W_j^k = p(z^k = 1 | w^j = 1) = \frac{p(w^j = 1 | z^k = 1) p(z^k = 1)}{\sum_{h=1}^K p(w^j = 1 | z^h = 1) p(z^h = 1)}. \tag{6}$$

As the topics are ancillary and unordered, we assume that the probability of a topic selection is uniformly distributed, i.e., $p(z^h = 1) = p(z^k = 1)$. It is conceivable that more sophisticated adaptive techniques for the probability of topic selection will result in a more accurate model, although we make a simpler uniform assumption in this paper. With this assumption, we obtain the probability of a word $w^j$ corresponding to a concept $z^k$ as

$$W_j^k = \frac{\beta_{jk}}{\sum_{h=1}^K \beta_{jh}}. \tag{7}$$

Furthermore, the documents can be represented in the topic space as well, $D_i \in \mathbb{R}^k$. The *kth* component $D_i^k$ of $D_i$ represents the probability of a concept $z^k$ given a document $\mathbf{d}_i$ and it is expressed as

$$\begin{aligned} D_i^k = p(z^k | \mathbf{d}_i) &= \sum_{w^j \in \mathbf{d}_i} p(z^k | w^j, \mathbf{d}_i) p(w^j | \mathbf{d}_i) \\ &= \sum_{w^j \in \mathbf{d}_i} p(z^k | w^j) p(w^j | \mathbf{d}_i), \end{aligned} \tag{8}$$

where we assume that the conditional probability $p(z^k | w^j, \mathbf{d}_i)$ equals the conditional probability $p(z^k | w^j)$. This assumption is based on the LDA assumption that there is a fixed number of underlying topics that are used to generate the words in documents (Croft et al., 2010). In other words, we assume that the words in topic space do not depend on which document it is used in, but on the topic it is generated from. An approximation $\tilde{D}_i^k$ of $D_i^k$ can be obtained by substituting $p(w^j | \mathbf{d}_i)$ with $\tilde{p}(w^j | \mathbf{d}_i)$, where

$$\tilde{p}(w^j | \mathbf{d}_i) = \frac{n_{ij}}{N_{\mathbf{d}_i}}. \tag{9}$$

Here, $n_{ij}$ denotes the number of occurrences of word $w^j$ in document $\mathbf{d}_i$ and $N_{\mathbf{d}_i}$ denotes the number of words in document $\mathbf{d}_i$, i.e., $N_{\mathbf{d}_i} = \sum_{j=1}^V n_{ij}$. We note that there could be many smoothing schemes to estimate $p(w^j | \mathbf{d}_i)$ although we have chosen the simplest form in this paper. Then,

$$\begin{aligned} D_i^k \simeq \tilde{D}_i^k &= \frac{\sum_{w^j \in \mathbf{d}_i} p(z^k | w^j) n_{ij}}{N_{\mathbf{d}_i}} \\ &= \frac{\sum_{w^j \in \mathbf{d}_i} W_j^k n_{ij}}{N_{\mathbf{d}_i}}. \end{aligned} \tag{10}$$

In general, a document includes various words that are used to explain key topics in the document. The definition of document probability in equation (10) captures the topical features of words in the document. This definition is distinguished from the usual definition of the probability of a document in LDA, which assumes that document probability is the same as the probability of the simultaneous occurrence of all words used in the document. The new definition overcomes the difficulty that is associated with the latter definition in which the probability of a document considerably depends on the length of the document. If topics are regarded as index terms, document representation in the topic space can be utilized for automatic document indexing. We call this novel indexing method LDI.

*Similarity between Document and Query*

One of the main applications of automatic indexing is document retrieval, which is also the main concern for LDI in this paper. With the new definitions in the previous subsection, each term can be represented in topic space, i.e., $W_j^k = \{W_j^1, W_j^2, ..., W_j^K\}$. Divided by norm, each term is unified to a unit circle. We can define the similarity $\rho(\cdot, \cdot)$ between two terms $w^s$ and $w^t$ as

$$\begin{aligned}\rho(w^s, w^t) &= \vec{W}_s \cdot \vec{W}_t \\ &= \sum_{k=1}^{K} \frac{p(z^k|w^s)p(z^k|w^t)}{\sum_{h=1}^{K} p(z^h|w^s) \sum_{h=1}^{K} p(z^h|w^t)} \\ &= \sum_{k=1}^{K} \frac{\beta_{sk}\beta_{tk}}{\sum_{h=1}^{K} \beta_{sh} \sum_{h=1}^{K} \beta_{th}},\end{aligned} \qquad (11)$$

where $\vec{W}_s = \frac{W_s}{\|W_s\|}$, $\vec{W}_t = \frac{W_t}{\|W_t\|}$, and $\vec{W}_s \cdot \vec{W}_t = \left\langle \frac{W_s}{\|W_s\|}, \frac{W_t}{\|W_t\|} \right\rangle$.

Hereafter, the capital letters represent probability vectors. The above similarity measure quantifies the proximity of two terms in topic space in terms of the cosine value of the angle between them. Thus, in general, the similarity between two distinct terms in general does not equal to zero, which mitigates the problem of synonymy. On the other hand, the problem of polysemy can also be alleviated since each term has multiple topical interpretations owing to the representation in a topic space.

Analogously, similarity measures to compare two documents and to compare a term and a document can be defined as

$$\rho(\mathbf{d}_s, \mathbf{d}_t) = \vec{D}_s \cdot \vec{D}_t \simeq \vec{\bar{D}}_s \cdot \vec{\bar{D}}_t \qquad (12)$$

and

$$\rho(w^s, \mathbf{d}_t) = \vec{W}_s \cdot \vec{D}_t \simeq \vec{W}_s \cdot \vec{\bar{D}}_t, \qquad (13)$$

respectively.

In terms of our original problem, we regard the query as a pseudo-document that contains a set of query terms $Q = \{q_1, q_2, ..., q_L\}$. Then, similar to (10), the probability vector of the query with respect to the *kth* topic can be defined in the concept space as

$$\begin{aligned}Q^k = p(z^k|Q) &= \sum_{q_j \in Q} p(z^k|q_j, Q)p(q_j|Q) \\ &\simeq \frac{\sum_{q_j \in Q} p(z^k|q_j)}{L}.\end{aligned} \qquad (14)$$

We note that $\sum_{q_j \in Q} p(z^k|q_j, Q)$ can be simplified as $\sum_{q_j \in Q} p(z^k|q_j)$. Similarity between query $Q$ and document $\mathbf{d}_s$ is measured by

$$\rho(\mathbf{d}_s, Q) = \vec{D}_s \cdot \vec{Q} \simeq \vec{\tilde{D}}_s \cdot \vec{\tilde{Q}}, \qquad (15)$$

where $\vec{Q} = \{Q^1, Q^2, ..., Q^K\}$ in the concept space.

Characterization of a query as a probability vector is possible because of the new definition of the document probability vector in (10). The probability vector represents the characteristics of the words, documents, and queries in the concept space. One advantage of LDI is that an unseen training query can be treated coherently as a document in the training set. This feature is pertinent to LDA, and it is not present in other automatic indexing methods such as pLSI (Hofmann, 2001).

The size of concept space $K$ plays an important role in our approach as it is rooted in LDA. In LDA, the value $K$ determines the degree of abstraction of information. The larger the value $K$ is, the finer is the segmentation of information.

*A Toy Example*

An example is considered for illustration.

> T1: the OS in Apple smartphones
> T2: the OS system in Apple products
> T3: the sign system in Samsung smartphones
> B1: Samsung and Apple signed a contract
> B2: there are many kinds of product contracts
> D1: fry the apple pie with some peas
> D2: the pie should be fried in oil
> D3: the way to fry dumplings
> G1: the oil is made from genetically-modified bean
> G2: the bean is genetically-modified from peas

As seen above, the example is made up of 10 documents with 14 different terms in four different disciplines: *technology*, *business*, *diet*, and *genetics*. Each document is labeled by the leading character of the related field, viz. T, B, D, and G, followed by the order of its appearance. For instance, G2 represents the second document in the *genetics* field. The term frequency is summarized in Table 1.

Figure 1 and Table 2 show the results of LDA applied to the example data with dimension $K = 4$. In both figures, the concepts, which would have not been known a priori, were ordered by *technology*, *business*, *diet*, and *genetics*. In Figure 1, as the term "*apple*" appears in more than one discipline, i.e., *technology*, *business*, and *diet*, the probability of the term representing each of these concepts spreads around the topics. With the new definition $W_j^k$ in (7), the term "*apple*" represents the concepts *technology* with probability 0.4275; *business*, 0.3053; *diet*, 0.2672; and *genetics*, 0. Other terms, such as "*smartphone*", "*contract*", "*pie*", "*genetically-modified*", appearing in one discipline lead to the probability of 1 in the respective topics.

Table 1: Summary of the characteristics of the toy example.

| Term | T1 | T2 | T3 | B1 | B2 | D1 | D2 | D3 | G1 | G2 |
|---|---|---|---|---|---|---|---|---|---|---|
| *apple* | 1 | 1 | | 1 | | 1 | | | | |
| *smartphone* | 1 | | 1 | | | | | | | |
| *os* | 1 | 1 | | | | | | | | |
| *system* | | 1 | 1 | | | | | | | |
| *product* | | 1 | | | 1 | | | | | |
| *contract* | | | | 1 | 1 | | | | | |
| *sign* | | | 1 | 1 | | | | | | |
| *samsung* | | | 1 | 1 | | | | | | |
| *pie* | | | | | | 1 | 1 | | | |
| *pea* | | | | | | 1 | | | | 1 |
| *fry* | | | | | | 1 | 1 | 1 | | |
| *oil* | | | | | | | 1 | | 1 | |
| *genetically-modified* | | | | | | | | | 1 | 1 |
| *bean* | | | | | | | | | 1 | 1 |

Table 2 exhibits an accurate assignment of document probabilities with respect to 4 topics based on the definition of $D_i^k$ in equation (10). Figure 2 shows the distribution of three arbitrary queries with respect to four topics, where the queries are given below. From the histogram in Figure 2, we observe that all three queries have no relationship with the fourth topic, *genetics*, and the definition $Q_i^k$ in (14) properly identified the relationship.

*Query 1: the sign system in Apple*
*Query 2: the contract for Apple products*
*Query 3: the Apple OS*

Table 3 shows the similarity between each query and each document, obtained using equation (15). Figure 3 is a plot of documents and queries on the first three topics space, which illustrates the similarity between the queries and documents. Obviously, documents T1, T2, and T3 are closely related to Query 1 and Query 3, and documents B1, B2 are closely related to Query 2. We note that the values in Table 3 do not possess the meaning in terms of probability although they are bounded above by 1 and below by 0. As for Query 3, although no term matches in document T3, the relevance between them is high because of the large probabilities in the same topic of *technology*. On the other hand, although the query term "*apple*" matches between Query 3 and D1, the similarity between them is calculated to be much smaller than that between Query 3 and T3 by the proposed method because of polysemy in the term "*apple*". The illustration in Figure 3 also verifies the significance of the cosine similarity. These are accurate characteristics of topic search that are not exhibited in keyword-based models. Although this particular example shows the effectiveness of the proposed model, the performance depends on training data and the assumption of latent topics.

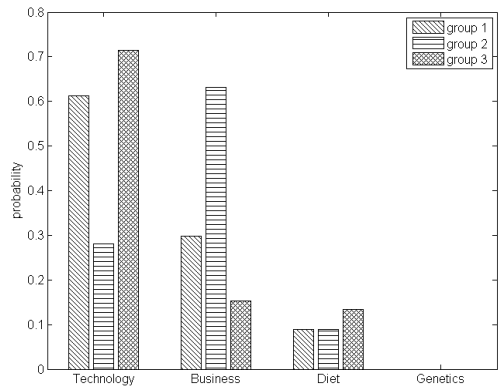

*Figure 1.* Assignment of $p(z|w)$ by $W_j^k$, where group 1={"*apple*"}, group 2 = {"*product*", "*sign*"} , and group 3 = {"*pea*", "*oil*"}.

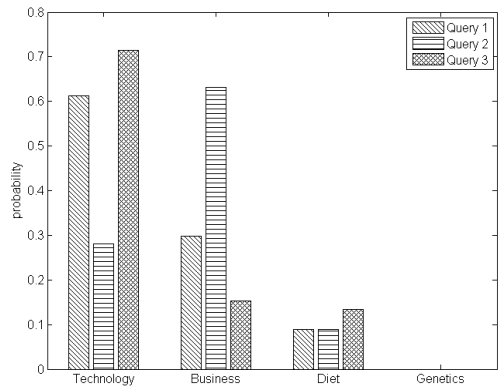

*Figure 2.* Distribution of query vectors in the concept space with four dimensions.

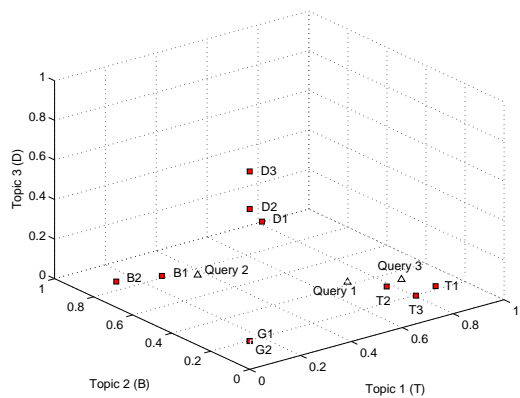

*Figure 3.* Queries and Documents in a three-dimensional topic space.

Table 2: Assignment of $p(z|d)$ by $D_i^k$.

| Document | Topic 1 | Topic 2 | Topic 3 | Topic 4 |
|---|---|---|---|---|
| T1 | **0.8092** | 0.1018 | 0.0891 | 0 |
| T2 | **0.7098** | 0.2234 | 0.0668 | 0 |
| T3 | **0.8039** | 0.1961 | 0 | 0 |
| B1 | 0.2098 | **0.7234** | 0.0668 | 0 |
| B2 | 0.1373 | **0.8627** | 0 | 0 |
| D1 | 0.1069 | 0.0763 | **0.6739** | 0.1429 |
| D2 | 0 | 0 | **0.8095** | 0.1905 |
| D3 | 0 | 0 | **1** | 0 |
| G1 | 0 | 0 | 0.1429 | **0.8571** |
| G2 | 0 | 0 | 0.1429 | **0.8571** |

Table 3: Similarities between queries and documents.

|  | Query 1 | Query 2 | Query 3 |
|---|---|---|---|
| T1 | **0.9475** | 0.5227 | **0.9938** |
| T2 | **0.9885** | 0.6643 | **0.9915** |
| T3 | **0.9692** | 0.6053 | **0.9833** |
| B1 | 0.6734 | **0.9902** | 0.4796 |
| B2 | 0.5681 | **0.9586** | 0.3543 |
| D1 | 0.3076 | 0.2829 | 0.3420 |
| D2 | 0.1261 | 0.1245 | 0.1752 |
| D3 | 0.1296 | 0.1279 | 0.1800 |
| G1 | 0.0213 | 0.0210 | 0.0296 |
| G2 | 0.0213 | 0.0210 | 0.0296 |

### Ensemble model

An EnM linearly combining indexing models is studied in this section. Different weights assigned to these constituent models may result in distinct results. We aim to derive a learning algorithm to obtain the optimal weights with which the EnM maximizes the MAP.

*Notations and Algorithm*

Suppose that the relevant document list for query $q_i \in Q$, is denoted as $D_i$. $|Q|$ represents the number of queries in the query set, and $|D_i|$, the number of documents in the relevant document set with respect to the query $q_i$. Let $d_{ij} \in D_i$ denote the $j$th document in $D_i$. A constituent document retrieval model $\phi_k$, chosen from a set of methods $\phi_k \in \Phi, \Phi = \{\phi_1, \phi_2, ..., \phi_{K_\phi}\}$, results in the similarity score denoted by $\phi_{ki}$ with respect to the query $q_i$. Accordingly, let $R(d_{ij}, \phi_{ki})$ be the ranking position of the $jth$ document for the $ith$ query returned by the $kth$ model. In order to measure the performance of the retrieved ranking, we use the widely used information retrieval metric, average precision (AP), which is denoted as $AP(\phi_{ki})$ associated with the $ith$ query and the $kth$ model given relevant document set $D_i$. Furthermore, we denote the metric, MAP, as $E(\phi_k)$ over the query set $Q$ calculated by the $kth$ model. The EnM is written as a linear combination of the constituent models, i.e., $H = \sum_{k=1}^{K_\phi} \alpha_k \phi_k$, by which the MAP is $E(H(\alpha))$, or $E(H)$ for simplicity. We assume that the relevant document list is sorted in descending order according to the scores. The definitions of AP and MAP are given below.

The AP of the $kth$ model $\phi_k$ given the $ith$ query $q_i$ is defined as

$$AP(\phi_{ki}) = \frac{1}{|D_i|} \sum_{j=1}^{|D_i|} \frac{j}{R(d_{ij}, \phi_{ki})}. \tag{16}$$

In this function, all documents are retrieved and the ranking positions of all relevant documents are used to define the precision. This metric is commonly used in literature (Gao, Qi, Xia, & Nie, 2005; Yue, Finley, Radlinski, & Joachims, 2007). Further, this metric can be used in VSM and VSM-based variants since relevant documents that contain no query terms are ranked low. The MAP of the $kth$ model $\phi_k$ for the query set $Q$ is defined as

$$E(\phi_k) = \frac{1}{|Q|} \sum_{i=1}^{|Q|} AP(\phi_{ki}). \tag{17}$$

As for the EnM $H$, MAP is defined as

$$\begin{aligned} E(H) &= \frac{1}{|Q|} \sum_{i=1}^{|Q|} AP(h_i(\alpha)) \\ &= \frac{1}{|Q|} \sum_{i=1}^{|Q|} \frac{1}{|D_i|} \sum_{j=1}^{|D_i|} \frac{j}{R(d_{ij}, h_i(\alpha))}. \end{aligned} \tag{18}$$

where $h_i(\alpha) = \sum_{k=1}^{K_\phi} \alpha_k \phi_{ki}$ is the similarity score returned by the EnM with weight vector $\alpha = (\alpha_1, ..., \alpha_{K_\phi})$ for the query $q_i$. Our goal is to find weights $\alpha_k$'s assigned to $\phi_k$'s with which the EnM gives the maximal MAP. Mathematically, we aim to solve the following problem

$$\max E(H). \tag{19}$$

Since the AP is in the range of $[0, 1]$, we can define the loss function of the above objective function as

$$Loss = \sum_{i=1}^{|Q|} (1 - AP(h_i(\alpha))). \tag{20}$$

In doing so, the maximization problem (19) is equivalent to the following minimization problem (21)

$$\min Loss. \tag{21}$$

According to the first-order Taylor series inequality $1 - x \leq e^{-x}$, we can instead minimize an upper bound of function (20)

$$Loss \leq \sum_{i=1}^{|Q|} \exp(-AP(h_i(\alpha))). \tag{22}$$

Equivalently, our aim is to minimize the following optimization problem

$$\min \sum_{i=1}^{|Q|} \exp(-AP(h_i(\alpha))). \tag{23}$$

where $AP(\cdot)$ is a nonconvex, nondifferentiable, noncontinuous function. A boosting-based algorithms is implemented to solve this problem.

Algorithm EnM.B is developed within the boosting scheme that utilizes the competition between constituent models and training data sets to iteratively update the weights until the game reaches equilibrium. To make the context consistent, we use the toy example introduced in the previous section and illustrate the operations of the algorithm in Figure 4. We assume that two rankers based on the *tf-idf* weight VSM (TFIDF) and the LDI are chosen as constituent methods and that the relevant document lists for Query 1, Query 2, and Query 3 are {T3}, {B1,B2}, and {T1, T2}, respectively. As shown in Figure 4, the query set is represented by the box on the left-hand side in which each circle denotes one query and the ranker set is denoted by the box on the right-hand side in which each circle denotes one ranker. The size of each circle in the box indicates the corresponding weight for each query and each ranker, whereas the font size of the term "Loss" indicates the value of the objective loss function.

---

**Algorithm** EnM.B: A boosting algorithm for training the EnM.

**Require:** Query set $Q$, a set of basis models $\Phi$, a duplicate set of models $\Phi' = \Phi$. Initialized weights $\mathfrak{D}^1$ over all queries with uniform distribution, i.e., $\mathfrak{D}^1 = 1/|Q|$. Initialize $\alpha$'s with zeros. Set the initial performance measure $E^0$.
  **while** $|E^t - E^{t-1}| > \varepsilon$ **do**
    **if** $\Phi' = \emptyset$ **then**
      $\Phi' = \Phi$;
    **end if**
    Select basis models $\phi^t \in \Phi'$ with weights $\mathfrak{D}^t$ on training queries using equation (37);
    Update the weight $\alpha = \alpha + \delta_{j^*} e_{j^*}$ using equation (35);
    Compute the MAP $E^t$ with EnM $H^t$;
    **if** $|E^t - E^{t-1}| > \varepsilon$ **then**
      $\Phi' = \Phi' \backslash \phi^t$;
      Update $\mathfrak{D}^{t+1}$ using (33);
    **end if**
  **end while**
  **return** EnM $H$.

---

In round 1, the EnM.B initializes uniform weights $\mathfrak{D}^1 = (1/3, 1/3, 1/3)$ for the queries and $\alpha = (0, 0)$ for the basis models. The constituent model TFIDF that minimizes the loss is chosen. Since the TFIDF performs well on some queries and poor on the other queries, the weight for the query on which TFIDF performs poorly is increased, which results in constructing weights $\mathfrak{D}^2 = (0.28, 0.43, 0.228)$ and $\alpha = (1.298, 0)$. From this step, we can observe that the weights $\mathfrak{D}^2$ play a penalty role on Query 2 so that the selection of the constituent model on round 2 focuses on Query 2. Then, the LDI model minimizing the loss is chosen in round 2. At the end of round 2, the weights are updated to $\mathfrak{D}^3 = (1/3, 1/3, 1/3)$ and $\alpha = (1.298, 1.285)$. The algorithm stops at round 3 since the loss converges to zero, i.e., the MAP equals to 1. We note that the original model set is reused until the algorithm converges. The convergence analysis is similar to the general boosting scheme, and thus, it is omitted in this article. The complexity of the EnM.B is $O(T \cdot K_\phi \cdot |Q| \cdot M)$, where $T$ indicates the number of iterations. The details of the derivations of this algorithm can be found in Appendix A.

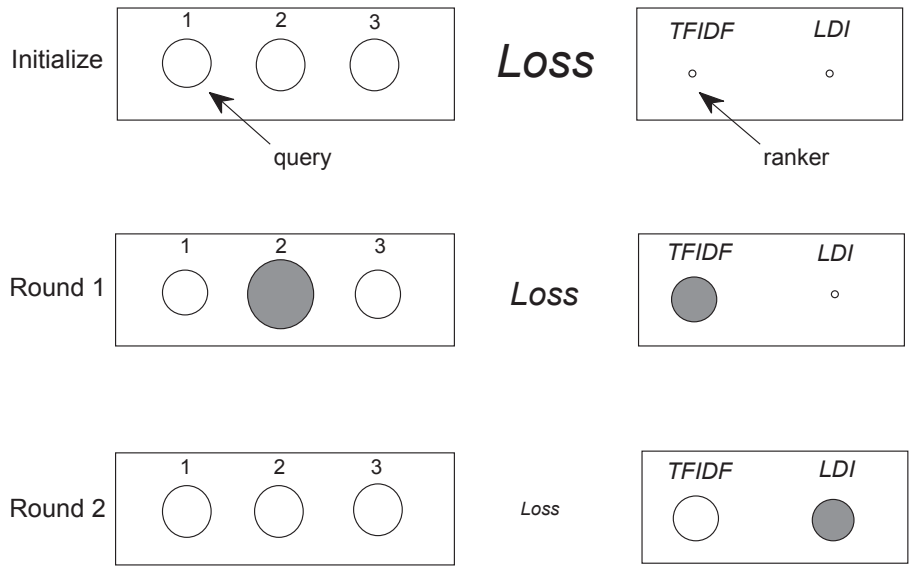

*Figure 4.* Account for how the EnM.B works with the toy example.

## Computational Experiments

*Experiment Setup*

Four standard data sets collected from four distinct subjects are utilized to test the proposed methods. They are provided in the SMART Information Retrieval System[1]. Each data set contains a corpus, a list of queries, and corresponding relevant documents. Since the data in each set are relatively small, we construct a Merged Collection (MC) that contains these four data sets. Although the size of MC is smaller than other large collections, such as the Text REtrieval Conference (TREC) Data, the experiment on MC can somewhat test the performances of the proposed methods on large data. Another reason of choosing the MC is because of the excessive computing time associated with LDA. To observe the structural information of the MC data set, we used t-distributed Stochastic Neighbor Embedding (t-SNE) (Van der Maaten & Hinton, 2008) to visualize the two-dimensional embedding of the proposed topical document representations in Figure 5. This drawing also reveals that the proposed document representations in topic space preserve essential structural information in the documents. Some basic data characteristics are summarized in Table 4.

---
[1] Available at: ftp://ftp.cs.cornell.edu/pub/smart.

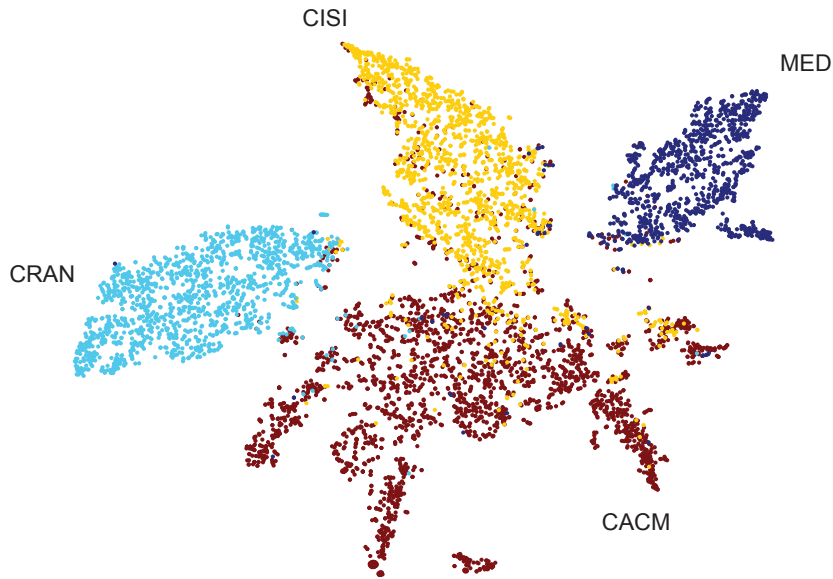

*Figure 5.* A two-dimensional embedding of the proposed document representations using t-SNE for the MC data set.

Prior to the proposed method, we applied the following simple preprocessing. Stop words were removed from all corpora using the list of 571 stop words provided in SMART. Special symbols, such as hyphenation marks, were removed and those words with a unique appearance in the corpus were also removed. Some documents and queries provided in CISI and CACM were vacant, which might decrease the retrieval accuracy. However, instead of removing those documents and queries, they were kept in the experiments to ensure the integrity of the data sets.

For comparison purpose in evaluating LDI, a *tf-idf* weight-based VSM model (denoted as TFIDF), LSI, pLSI, and LBDM were also tested on the same data sets. The TFIDF and LSI methods were coded, the LBDM was implemented based on GibbsLDA++[2], and the pLSA[3], and LDA[4] codes were obtained from the publicly available sites.

In the experiment for verifying EnM, TFIDF, LSI, pLSI, and LDI are used as the constituent indexing models. The advantage of using these four constituent models is that both the conceptual meaning and keyword matching information can be combined into the EnM. We believe that the EnM can benefit from the combined information and, in turn, result in an overall improvement in ranking accuracy. In order to address overfitting, each experimental data set is divided into two equivalent parts, and the EnM is evaluated through two-fold crossvalidation. A uniform weighted EnM (denoted as UniEnM) is also implemented for comparison.

---

[2]Available at: http://gibbslda.sourceforge.net/.
[3]Available at: http://www.kyb.mpg.de/bs/people/pgehler/code/index.html.
[4]Available at: http://www.cs.princeton.edu/ blei/lda-c/.

Table 4: Data characteristics.

| Data | Subject | Document # | Query # | Term # |
|------|---------|------------|---------|--------|
| MED  | M(Medicine)    | 1,033 | 30  | 5,775  |
| CRAN | A(Aeronautics) | 1,400 | 225 | 8,213  |
| CISI | L(Library)     | 1,460 | 112 | 10,170 |
| CACM | C(Computer)    | 3,204 | 64  | 9,961  |
| MC   | M, A, L and C  | 7,097 | 431 | 27,784 |

Table 5: Number of topics used in LSI, pLSI and LDA.

| Method | MED | CRAN | CISI | CACM | MC  |
|--------|-----|------|------|------|-----|
| LSI    | 100 | 125  | 150  | 125  | 500 |
| pLSI   | 100 | 150  | 50   | 75   | 400 |
| LDI    | 100 | 100  | 100  | 75   | 200 |

*Parameter Selection*

Choosing an appropriate number of topics is an important issue for LSI, pLSI, and LDI. In this paper, we followed the strategy suggested by (Deerwester et al., 1990). That is, we examined the performance for different numbers of dimensions and selected the one that maximizes retrieval performance. Table 5 summarizes the number of dimensions for each method on different data sets. Validations for these selections can be found in Appendix B. In applying pLSI, we modified the determination of parameter $\beta$. In the search of the $\beta$ value in the tempered EM in pLSI, the $\beta$ value was reduced until perplexity could no longer be reduced. However, in our experiment, we further reduced the $\beta$ value until the precision did not improve. Following the strategy given by X. Wei and Croft (2006) for LBDM, we set the number of topics identical to LDI, $\lambda = 0.7$ and $\mu = 1000$, which are the optimal settings.

*Experimental Results of LDI*

Figs. 6 and 7 depict the precision-recall curves of the tested methods on four standard data sets and MC, respectively. The numerical values of the x-axis denote the recall of labeled relevant documents, while the numerical values of the y-axis represent the precision of retrieved documents. Compared to TFIDF, LSI, pLSI, and LBDM, the proposed LDI chieved the best performance, except for small intervals in recalls on MED and CRAN. In the experiments on CRAN, CISI, and CACM, LDI had a higher precision for a high recall regime. This property seems more valuable in practice since the documents in higher positions are viewed by more users. As for the experiment on MC, the proposed LDI performed better than the indexing methods TFIDF, LSI, and pLSI, and showed a comparable performance with the retrieval model LBDM. In addition to the performance superiority to the LBDM, the proposed LDI also has advantages in the following two aspects: (1) the topical representation of a document for automatic indexing is proposed in our method, while it is not introduced by the LBDM; and (2) no smoothing parameters need to be tuned in our method whereas additional parameters need to be trained in the LBDM.

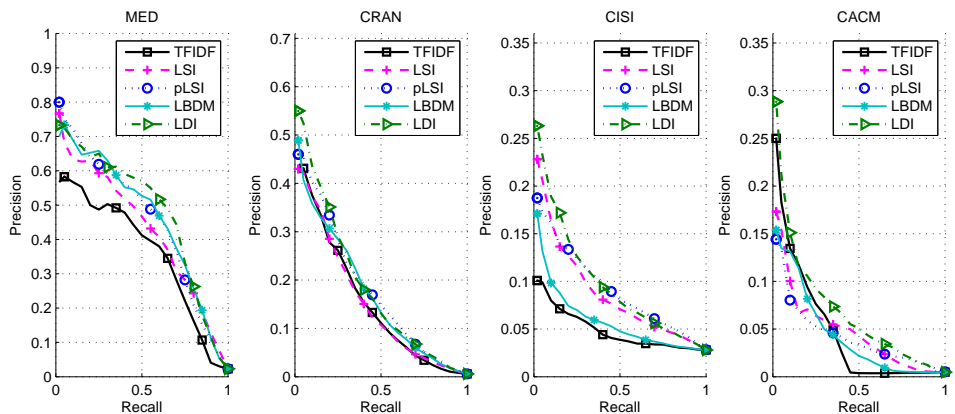

*Figure 6.* Precision-Recall Curves for TFIDF, LSI, pLSI, LBDM, and LDI on MED, CRAN, CISI, and CACM.

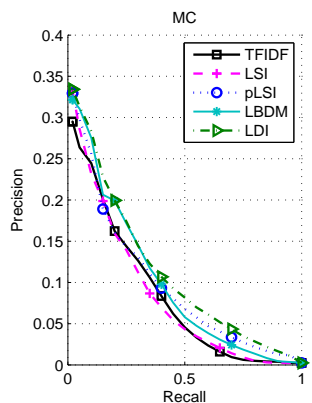

*Figure 7.* Precision-Recall Curves of TFIDF, LSI, pLSI, LBDM, and LDI on MC.

*Experimental Results of the EnM*

The aim of the first experiment was to examine the learning ability of EnM.B. Owing to length constraints, we consider the MED corpus as an example. The experiments on other corpora have similar results. The learning curve of EnM.B in terms of MAP are shown in Figure 8 during the trials of cross validation. From the figure, we can observe that the EnM.B converges as the number of rounds increases, which is consistent with our analysis.

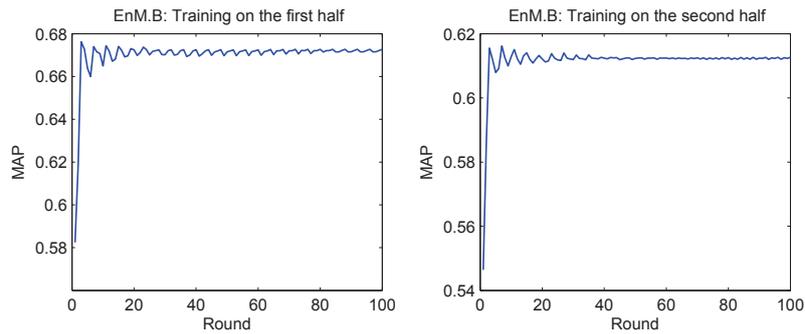

*Figure 8.* Learning curve of EnM.B on MED.

We then compute the weights for the constituent models. Since the MAP is determined by the relative ratios of the basis models, the normalization of the weights does not change the final MAP value. We normalized the mean values of weights over all trials of cross validation, as listed in Table 6.

Table 6: Normalized weights for the constituent indexing models in the EnM.

| Data | TFIDF | LSI | pLSI | LDI |
|------|-------|------|------|------|
| MED  | 0.21  | 0.21 | 0.25 | 0.33 |
| CRAN | 0.22  | 0.22 | 0.25 | 0.31 |
| CISI | 0.31  | 0.18 | 0.18 | 0.33 |
| CACM | 0.47  | 0.28 | 0.12 | 0.13 |
| MC   | 0.34  | 0.18 | 0.26 | 0.22 |

With the abovementioned corresponding weights, Figs. 9 and 10 illustrate the precision-recall curves of the EnM on four standard data sets and MC, respectively. Here, we use the optimal weights of EnM.B over all trials to represent the EnM. As expected, the performance of the EnM outperforms LDI since the EnM is supervised with the knowledge of relevant documents whereas LDI is unsupervised without training on relevant documents. The EnM is prior to UniEnM as well since the weights are updated to optimal in EnM.

As shown in Figs. 7 and 10, the performance of LDI is moderate when tested with the MC data set. This may be due to the limitation of the data sets in which the relevant documents are given within each specific corpus. In detail, given a query, a document in another corpus may appear to be intuitively relevant by general users. This document may be retrieved by LDI but not listed in the relevant list. Since this kind of relevant documents from other corpora are excluded in the expert identified relevant list, the precision of retrieving these documents is relatively low. Two examples in LDI are given below to show this phenomenon. Two documents retrieved at the first ranking position in CACM and CISI are seen to be intuitively relevant to the corresponding query in CISI and CACM, respectively. However, they are excluded in the relevant list. This is because some documents in CISI and CACM may be relevant because of common subjects, which can also be observed in Figure 5. Therefore, the performances of the proposed methods on large data sets with a precise relevant list still require further research.

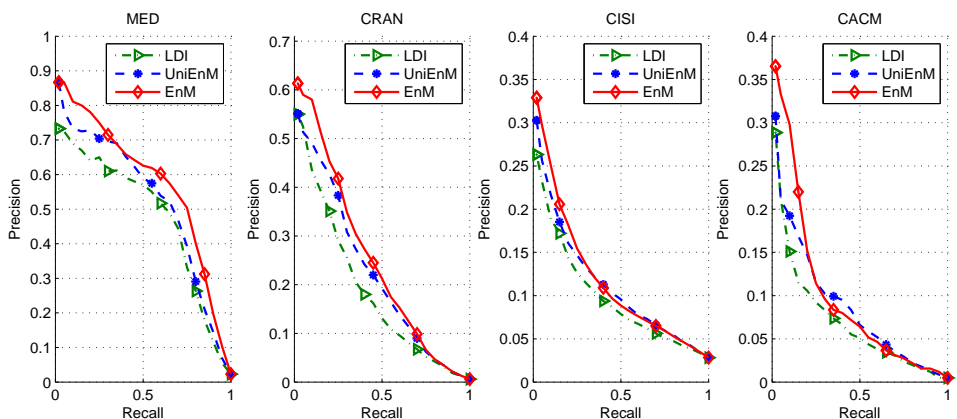

*Figure 9.* Precision-Recall Curves of LDI, UniEnM and EnM on MED, CRAN, CISI and CACM.

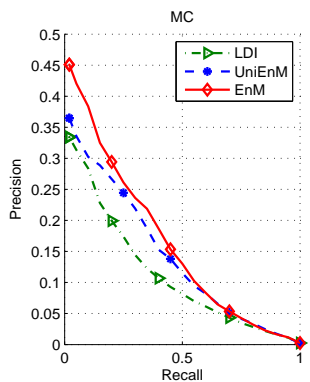

*Figure 10.* Precision-Recall Curves of LDI, UniEnM and EnM on MC.

**Example 1:**
*CISI: QueryID=6*
*What possibilities are there for verbal communication between computers and humans, that is, communication via the spoken word?*
*CACM: DocumentID=698, RankingPosition=1*
*DATA-DIAL: Two-Way Communication with Computers From Ordinary dial Telephones. An operating system is described which allows users to call up a remotely located computer from ordinary dial telephones. No special hardware or connections are required at the users' telephones. Input to the computer is through the telephone dial;output from the computer is in spoken form.*

Table 7: MAPs of various methods on the tested data sets. impr(%) indicates the percentage of improvement over TFIDF.

| Method | MED MAP | MED impr(%) | CRAN MAP | CRAN impr(%) | CISI MAP | CISI impr(%) | CACM MAP | CACM impr(%) | MC MAP | MC impr(%) |
|---|---|---|---|---|---|---|---|---|---|---|
| TFIDF | 0.4605 | - | 0.2716 | - | 0.0935 | - | 0.1177 | - | 0.1946 | - |
| LSI | 0.5026 | +9.1 | 0.2661 | -2.0 | 0.1229 | +24.0 | 0.1094 | -7.1 | 0.1709 | -12.2 |
| pLSI | 0.5334 | +15.8 | 0.2740 | +0.9 | 0.1223 | +23.4 | 0.0973 | -17.3 | 0.1918 | -1.4 |
| LBDM | 0.5516 | +19.8 | 0.2941 | +8.3 | 0.1030 | +10.2 | 0.1174 | -0.3 | 0.1968 | +1.1 |
| LDI | 0.5738 | +24.6 | 0.3146 | +15.8 | 0.1429 | +52.8 | 0.1349 | +14.6 | 0.2084 | +7.1 |
| UniEnM | 0.6034 | +31.0 | 0.3519 | +29.6 | 0.1510 | +61.5 | 0.1743 | +48.1 | 0.2735 | +40.5 |
| EnM | 0.6420 | +39.4 | 0.3766 | +38.7 | 0.1637 | +75.1 | 0.1890 | +60.6 | 0.2768 | +42.2 |

*Results of a test with telephones in the Boston area are reported.*
**Example 2:**
*CACM: QueryID=59*
*Dictionary construction and accessing methods for fast retrieval of words or lexical items or morphologically related information. Hashing or indexing methods are usually applied to English spelling or natural language problems.*
*CISI: DocumentID=179, RankingPosition=1*
*This book deals with the computer processing of large information files, with special emphasis on automatic text handling methods. Described in particular are procedures for dictionary construction and dictionary look-up, statistical and syntactic language analysis methods, information search and matching procedures, automatic information dissemination systems, and methods for user interaction with the mechanized system. As such, the text includes elements of linguistics, mathematics, and computer programming.*

Table 7 provides a summary of the overall performance of each testing model. The performance is measured by the MAP and the percentage of improvement over TFIDF. From this table, we can similarly conclude that the LDI achieves the highest precision among indexing and retrieval models, and EnM outperforms each constituent indexing model. This result verifies the general belief that the EnM performs better than individual models.

## Discussions and Conclusions

We presented a document indexing method, LDI, based on LDA. The proposed method utilized the $\beta$ matrix of LDA in defining document representations in topic space. The proposed novel document representation was subsequently used for retrieval by computing the similarity scores between the documents and the given query. The results of our computational experiments indicate that the proposed LDI method is a viable automatic DI method for information retrieval. Since LDI is a generative model, it can be employed in the information systems without labeled items, such as search engines, as to retrieve relevant documents according to the user input queries.

Another contribution lied in the proposition of the EnM. This model was inspired by the general belief that the combined model performs better than a single model. By formulating an optimization problem that minimized the loss defined based on the MAP, we proposed an algorithm, EnM.B, to solve this problem and obtain the optimal weights for each constituent model. The empirical results have shown that the EnM outperforms any constituent model through the overall recall regimes. Unlike LDI, the EnM can be only used in those systems with the knowledge of relevant documents, such as library systems.

In addition, two questions draw our interests for the EnM. (a) Does the selection of the constituent ranking models influence the result of EnM? (b) Are the results of the EnM global optimal to the optimization problem? Intuitively, the answer is yes for the first question because, empirically, the EnM benefits from the LDI and other basis models. The selection of constituent models and the number of constituent models are indeed important factors, which require further discussion. As for the second question, the answer is unknown. Though the algorithm EnM.B generates results and converges as iterations increase, no evidence shows that the final results are global optimal, or even local optimal. Thus, the proof of optimality requires further research. Besides, a precise algorithm for solving the optimization problem is also a subject that should be studied further. However, by and large, the EnM is still a considerable method for information retrieval.

## Acknowledgements

This work was partially supported by the National Research Foundation of Korea (NRF) grant funded by the Korea government (MEST) (2009-0083893). The first author gratefully acknowledges the China Scholarship Council (CSC) for fellowship support. We also acknowledge the referees and the editors for their suggestions and comments.

# Appendix A
# Derivation of the EnM.B

Suppose that at step $t$, the similarity score vector for the query $q_i$ returned by the $\alpha^t$ weighted EnM is

$$h_i^t(\alpha^t) = \sum_{k=1}^{K_\phi} \alpha_k^t \phi_{ki}, \qquad (24)$$

where $\alpha^t = (\alpha_1^t, ..., \alpha_{K_\phi}^t)$. Let $j^*$ and $\delta_{j^*}^t$ be chosen such that the loss is maximally reduced. The update of $\alpha^t$ at step $t+1$ can be written as $\alpha^{t+1} = \alpha^t + \delta_{j^*}^t e_{j^*}^t$ where $\delta_{j^*}^t$ is the step length and $e_{j^*}^t$ is the vector with 1 at the $j^*$ position and zero everywhere else. Our goal is to determine $j^*$ and $\delta_j^*$. The loss function at step *t+1* can be written as

$$L(\alpha_1, ..., \alpha_j + \delta_j, ..., \alpha_{K_\phi}) = \sum_{i=1}^{|Q|} \exp\left(-AP(h_i(\alpha^{t+1}))\right)$$

$$= \sum_{i=1}^{|Q|} \exp\left(-AP\left(\sum_{k=1}^{K_\phi} \alpha_k \phi_{ki} + \delta_j \phi_{ji}\right)\right) \quad (25)$$

$$= \sum_{i=1}^{|Q|} \exp\left(-AP(h_i(\alpha) + \delta_j \phi_{ji})\right).$$

For notational simplicity, we represent $L(\alpha_1, ..., \alpha_j + \delta_j, ..., \alpha_{K_\phi})$ by $L(\delta_j)$, $AP(h_i(\alpha))$ by $AP(h_i)$, and omit superscript $t$ henceforth.

Define
$$\Delta(h_i(\alpha^{t+1})) = AP(h_i + \delta_j \phi_{ji}) - AP(h_i) - \delta_j AP(\phi_{ji}). \quad (26)$$

Then equation (25) can be written as

$$L(\delta_j) = \sum_{i=1}^{|\mathbf{Q}|} \exp(-AP(h_i)) \exp(-\delta_j AP(\phi_{ji})) exp(-\Delta(h_i(\alpha^{t+1}))). \quad (27)$$

Since AP ranges from 0 to 1, we can get $\Delta(h_i(\alpha^{t+1})) \in [-1-\delta_j, 1]$ and $\exp(-\Delta(h_i(\alpha^{t+1}))) \in [e^{-1}, e^{1+\delta_j}]$. There always exits a sufficiently large constant $M \geq e^{1+\delta_j}$ so that the following inequality holds

$$L(\delta_j) \leq M \sum_{i=1}^{|Q|} \exp(-AP(h_i)) \exp\left(-\delta_j AP(\phi_{ji})\right). \quad (28)$$

According to the inequality

$$e^{\alpha x} \leq \frac{1+x}{2e^\alpha} + \frac{1-x}{2e^{-\alpha}}, \quad (29)$$

we have
$$L(\delta_j) \leq M J(\delta_j), \quad (30)$$

where
$$J(\delta_j) = \sum_{i=1}^{|Q|} \exp(-AP(h_i)) \left\{\frac{1 + AP(\phi_{ji})}{2} e^{-\delta_j} + \frac{1 - AP(\phi_{ji})}{2} e^{\delta_j}\right\}. \quad (31)$$

Since $J(\delta_j)$ is convex with respect to $\delta_j$, we can minimize it by setting

$$\frac{\partial J(\delta_j)}{\partial \delta_j} = 0. \quad (32)$$

Let
$$\mathfrak{D}_i = \frac{\exp(-AP(h_i))}{Z}, \quad (33)$$

where $Z = \sum_{i=1}^{|Q|} \exp(-AP(h_i))$ is a normalization factor in the current step. We can observe that $\mathfrak{D}$ is a proportion over queries to penalize those queries on which the EnM has poor performance. Then, equation (32) can be written as

$$-\frac{\sum_{i=1}^{|Q|} \mathfrak{D}_i \left(1 + AP(\phi_{ji})\right)}{2} e^{-\delta_j} + \frac{\sum_{i=1}^{|Q|} \mathfrak{D}_i \left(1 - AP(\phi_{ji})\right)}{2} e^{\delta_j} = 0. \qquad (34)$$

Solving the above equation, we get

$$\delta_j = \frac{1}{2} \log \frac{\sum_{i=1}^{|Q|} \mathfrak{D}_i \left(1 + AP(\phi_{ji})\right)}{\sum_{i=1}^{|Q|} \mathfrak{D}_i \left(1 - AP(\phi_{ji})\right)} \qquad (35)$$

Plugging (35) in (31), we can choose the $j^*$ that minimizes the objective function

$$j^* = \arg\min_j \sum_{i=1}^{|Q|} \mathfrak{D}_i \sqrt{\left(1 + AP(\phi_{ji})\right)\left(1 - AP(\phi_{ji})\right)}, \qquad (36)$$

which can be simplified to

$$j^* = \arg\max_j \sum_{i=1}^{|Q|} \mathfrak{D}_i AP(\phi_{ji}). \qquad (37)$$

## Appendix B
## Validation of the selection of number of topics

In the experiment, we explored various number of topics from 50 to 150 for MED, CRAN, CISI, and CACM at a step size of 25 and from 100 to 500 for MC at a step size of 100. The following tables list most of the results in terms of MAP. For LSI, we used a small number of topics for computational simplicity, as suggested by Deerwester et al. (1990), though the performance might slightly improve if a larger number of topics were explored. As for pLSI and LDI, we observed a more clearly identifiable maxima, i.e., the performance increased up to a certain point and then decreased.

Table B1: MAP with different number of topics on MED.

| Number of topics | LSI | pLSI | LDI |
|---|---|---|---|
| $K = 50$ | 0.4538 | 0.4534 | 0.4868 |
| $K = 75$ | 0.4885 | 0.4448 | 0.5615 |
| $K = 100$ | **0.5026** | **0.5334** | **0.5738** |
| $K = 125$ | 0.4976 | 0.4089 | 0.5687 |

Table B2: MAP with different number of topics on CRAN.

| Number of topics | LSI | pLSI | LDI |
|---|---|---|---|
| $K = 50$ | 0.2151 | 0.2495 | 0.2767 |
| $K = 75$ | 0.2383 | 0.2464 | 0.2931 |
| $K = 100$ | 0.2541 | 0.2606 | **0.3146** |
| $K = 125$ | **0.2661** | 0.2681 | 0.3133 |
| $K = 150$ | 0.2668 | **0.2740** | 0.3002 |

Table B3: MAP with different number of topics on CISI.

| Number of topics | LSI | pLSI | LDI |
|---|---|---|---|
| $K = 50$ | 0.1086 | **0.1223** | 0.1124 |
| $K = 75$ | 0.1139 | 0.1062 | 0.1188 |
| $K = 100$ | 0.1185 | 0.1093 | **0.1429** |
| $K = 125$ | 0.1210 | 0.1079 | 0.1413 |
| $K = 150$ | **0.1229** | 0.1056 | 0.1378 |

Table B4: MAP with different number of topics on CACM.

| Number of topics | LSI | pLSI | LDI |
|---|---|---|---|
| $K = 50$ | 0.0601 | 0.0703 | 0.0908 |
| $K = 75$ | 0.0817 | **0.0973** | **0.1349** |
| $K = 100$ | 0.0962 | 0.0920 | 0.1083 |
| $K = 125$ | **0.1094** | 0.0776 | 0.1256 |

Table B5: MAP with different number of topics on MC.

| Number of topics | LSI | pLSI | LDI |
|---|---|---|---|
| $K = 100$ | 0.1150 | 0.1286 | 0.1906 |
| $K = 150$ | 0.1339 | 0.1119 | 0.1973 |
| $K = 200$ | 0.1450 | 0.1194 | **0.2084** |
| $K = 250$ | 0.1541 | 0.1830 | 0.2011 |
| $K = 400$ | 0.1681 | **0.1918** | 0.1908 |
| $K = 500$ | **0.1709** | 0.1909 | 0.1871 |